# MAGNETOSTIMULATED CHANGES OF MICROHARDNESS IN POTASSIUM ACID PHTHALATE CRYSTALS

**M. V. Koldaeva, T. N. Turskaya, and E. V. Darinskaya**

*Shubnikov Institute of Crystallography, Russian Academy of Sciences,*
*Leninskii pr. 59, Moscow, 119333 Russia*
*e-mail: mkoldaeva@ns.crys.ras.ru*

A decrease in microhardness along the (010) cleavage in potassium acid phthalate single crystals by 15–18% after the application of a permanent magnetic field is revealed for the first time.[1] It is shown that the effect revealed is of the volume character. The role of interlayer water in the processes stimulated by a magnetic field is studied. Interlayer water does not cause the observed changes; it only plays the part of an indicator of these changes in potassium acid phthalate crystals in a magnetic field. It is established that microhardness in the (100) plane of the crystal in an applied a magnetic field first increases by 12–15% and then remains constant in time within the accuracy of the experiment. The possibility of varying the crystal structure of potassium acid phthalate crystals by applying magnetic fields inducing rearrangement in the system of hydrogen bonds or in the defect structure is discussed.

The influence of weak magnetic fields on the mechanical properties of nonmagnetic crystals (magnetoplastic effect) is observed in both micro- [1] and macroplasticity [2–9] and is studied by scientists who use plastic physics methods.

The studies started in 1987 [1] showed that the magnetoplastic effect is explained by the action of a magnetic field on spin-dependent electron transitions either in a system dislocation–paramagnetic center [10, 11] or in magnetosensitive complexes of point defects [12, 13]. The electron transitions stimulated by magnetic fields may change the local energy of the dislocation interaction with a point defect, which results in plasticization (or hardening [14]) of a number of diamagnetic alkali halide, semiconductor, and metal crystals [10, 11]. As was shown earlier, similar processes may also change the rates of chemical reactions in magnetic fields [15, 16]. Up to now, the study of the magnetoplastic effect has been performed mainly on isotropic crystals. Interest has arisen in the study of the influence of a magnetic field on the properties of nonmagnetic anisotropic crystals with complicated structures widely used in technology.

We studied a potassium acid phthalate (KAP) crystal of the composition $C_8H_5O_4K$ with ionic, covalent, and hydrogen bonds [17, 18]. An orthorhombic crystal is described by the point symmetry group *mm*2. The crystal properties are explained by the presence of the polar 001 axis and a large distance between the (010) cleavages (~13 Å). Figure 1 shows the structural formula of KAP crystals possessing piezoelectric properties [19] that are widely used as analyzers in the long-wavelength range of the X-ray spectrum and as monochromators in various high-resolution X-ray instruments [20]. In the temperature range from 300 to 2 K, KAP crystals are diamagnetic along all the three main crystallographic directions.[2]

## SAMPLE PREPARATION AND PROCESSING OF RESULTS

The experimental methods used in our study were selected in accordance with the studies of mechanical properties of KAP crystals [21–23] and, in particular, of their microhardness [23]. Motion of dislocations was not studied in any cited works. In [21, 22], considerable anisotropy of crystal deformation was observed: depending on the mutual orientation of the compression and polar (*z*) axes, crystals demonstrated brittle fracture, kink bands, or plastic deformation. To detect the influence of a magnetic field on the crystals we used a fast technically simple method of microindentation.

The KAP crystals used in our experiments were grown from aqueous solution by the method of decreasing temperature [19] at the Institute of Crystallography of the Russian Academy of Sciences. The samples were cut from crystalline boules by a wet thread. The faces were mechanically polished against a smooth wet silk. Microhardness was measured in the (010) and (100) faces. To measure microhardness in the (010) plane, we cleaved thin 3 × 0.3 × 7 mm large plates along the cleavage. The 3 × 3 × 7 mm samples for measurements in the (100) face were additionally polished prior to the experiment first mechanically and then by chemical polishing to remove a thin (about 0.15 mm) surface layer. Indentation was made with a Vickers pyramid under a load of 0.2 N, the impression diagonal was measured with the aid of a standard attachment to a Neophot-21 microscope.

---

[1] Most of the results obtained in this study were first preseted at the First Russian Conference of Young Scientists on Physical Material Science in Kaluga, October 4–7, 2001 (M.V. Koldaeva, *Collected Abstracts* (Kaluga, Manuskript, 2001), pp. 39–40).

[2] Magnetic susceptibility of KAP crystals was measured by Yu.G. Shvedenkov at the International Tomography Center, Siberian Division, Russian Academy of Sciences, Novosibirsk.

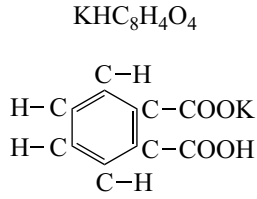

**Fig. 1.** Structural formula of a potassium acid phthalate (KAP) crystal.

Figure 2 (courtesy of N.L. Sizova from the Institute of Crystallography) shows typical impressions in the (010) cleavage (Fig. 2a) and in the perpendicular (100) plane (Fig. 2b) with the singled-out polar direction $z$ and the diagonal $d$. Figure 2a shows that the impression in the (010) cleavage gives rise to cracking characteristic of brittle crystals [24]. Nevertheless the impression has an obvious faceting which allows one to make measurements. To measure microhardness in the (100) plane, the crystal was oriented in such a way that the angle formed by $d$ and $z$ was about 45°. The impression was almost a square shape; no cracks were recorded (Fig. 2b).

To avoid any confusion associated with the uniqueness of each measured surface and possible dependence of the results obtained on atmospheric conditions, each experimental point was measured on 3–5 thin plates sawed from two different bulky samples cut from the same growth pyramid. The microhardness of these plates was the same. Figure 3 shows a typical histogram of the diagonal $d$ of the impression measured on face (010) of different samples. It is seen that the histogram is characterized by a normal Gaussian distribution. Thus, physically, averaging along the impression diagonals $d$ measured under the same conditions on different cleavages is quite justified. The statistical errors $\Delta d$ were calculated using 50–100 measurements by the Microsoft Excel program and ranged within 2–4%. Microhardness $H$ (in GPa) was calculated by the conventional formula for a Vickers pyramid [24] as

$$H = 1.854 \times 10^5 P/d_{st}^2,$$

where $d_{st}$ is the statistical average of the impression diagonals measured in microns and $P$ is the load applied to an indenter in grams. The measurement error $\Delta H$ was calculated as

$$\Delta H = 2H(\Delta d/d_{st}),$$

where $\Delta d$ is the statistical error of diagonal averaging over the sampling.

## EXPERIMENTAL RESULTS AND DISCUSSION

In the first run of experiments, the samples in the shape of thin 3 × 7 mm plates with a thickness of ~0.3 mm were cleaved along the cleavage plane of one of the growth pyramids. One of two as-cleaved mirror-smooth surfaces was placed for 5 min into a 0.9 T magnetic field (**B** ∥ $z$), whereas the other sample served as a reference. Then microhardness of the surfaces of both samples was measured for 7–10 days as a function of time passed since cleavage. The results of these experiments are shown in Fig. 4a. After keeping the sample for five minutes in a 0.9 T magnetic field, microhardness decreased by 15–18% (curve 2) in comparison with the microhardness of the mirror-smooth cleavage of the reference sample (curve 1). This difference decreased with time and disappeared after 7–8 days. It was also interesting to consider microhardness as a function of time passed since the sample preparation. Therefore, we studied the surface of the reference sample (Fig. 4a, curve 1). The removal of interlayer water from an as-cleaved surface was studied in [25]. Inter-layer water is always present in potassium acid phthalate crystals but it is not a structure-forming element. Possibly softening of the cleavage with time is explained by such processes.

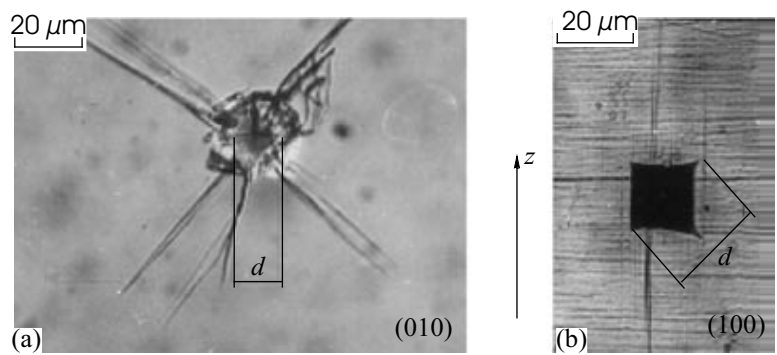

**Fig. 2.** Typical impression of an indenter with the single-out $z$ direction and the diagonal $d$ (a) in the (010) and (b) (100) planes.

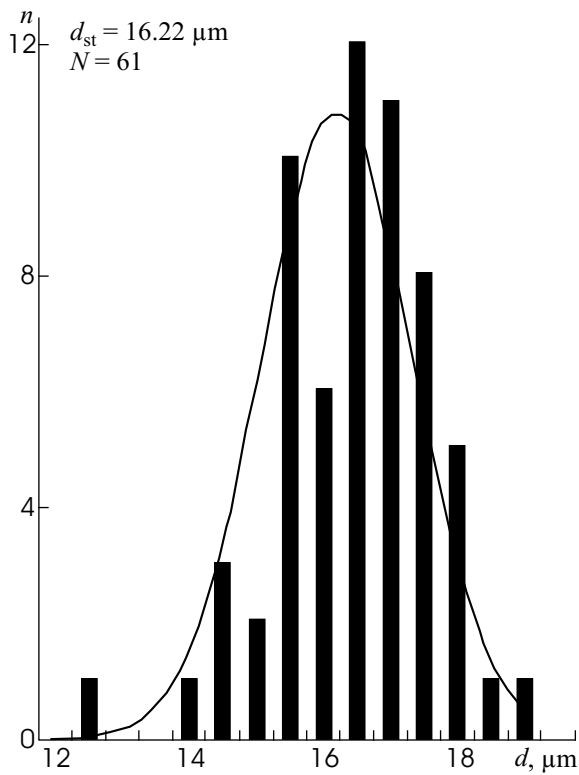

**Fig. 3.** Histogram of diagonals $d$ of indenter impressions for four samples; the solid curve is the normal distribution of $d$ calculated from the average $d_{st}$ value and dispersion over the histogram.

To clarify the role of interlayer water in magnetostimulated processes, we had to diminish the amount of interlayer water in the bulk of potassium acid phthalate crystals. With this aim, we annealed the crystals for six hour at 220°C and then slowly cooled the crystals in an argon flow. With an increase of the temperature, water located in interplanar space leaves the crystal but is absorbed again during subsequent cooling [25]. To prevent water absorption, the furnace was constantly blown with a flow of argon both during annealing and cooling. The magnetostimulated changes of microhardness in annealed crystals were studied according the same scheme as in unannealed crystals.

The microhardness curve of the as-cleaved surfaces of annealed reference samples (Fig. 4b, curve *1*) is considerably lower than microhardness curve of unannealed samples (Fig. 4a, curve *1*). However, in this case as well, microhardness of annealed samples treated in a magnetic field (curve *2*) decreased by 15–18% in comparison with microhardness of the annealed reference samples.

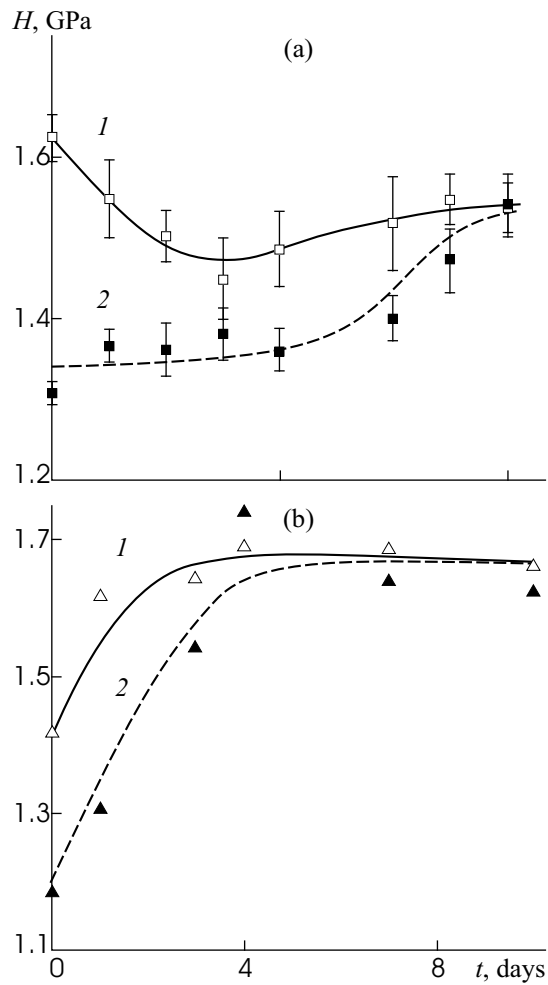

**Fig. 4.** Microhardness of the (010) face of a potassium acid phthalate crystal as a function of time passed since the preparation of mirror-smooth cleavages: (*1*) reference sample, (*2*) sample treated for 5 min in a 0.9 T magnetic field, (a) unannealed and (b) annealed crystals. In all the cases B ∥ z.

Comparing the kinetics of surface aging (which manifests itself in the changes of microhardness with time) in unannealed and annealed samples (curves *1* in Figs. 4a, 4b, curves *1*), we assumed that surface aging of potassium acid phthalate crystals is associated with loss of water after preparation of cleavage and the interaction of its surface with atmosphere. After the treatment in a magnetic field, surface aging of unannealed crystals proceeds much slower than in the samples not kept in a magnetic field (Fig. 4a). At the same time, the initial slope of the $H(t)$ curves of the annealed crystals containing much less interlayer water did not change after treatment in a magnetic field (Fig. 4b). Thus, interlayer water is only an indicator of the changes taking place in potassium acid phthalate crystals in a magnetic crystal, and it cannot give rise to these changes.

To establish whether it is only a surface effect or the magnetic field that gives rise to the changes in the crystal bulk, we broke a bulky sample into two parts. Like in the first run of experiments, one part of the crystal was kept for five min in a 0.9 T magnetic field (**B** ∥ *z*), whereas the other part served as the reference sample. In this case, before every measurement a new cleavage was made, which allowed us to follow the possible changes in the bulk of the crystal subjected to the action of a magnetic field. To improve statistics, three bulk samples were used in these experiments.

Within the accuracy of the experiment, no temporal changes were observed in the reference part of the crystal (Fig. 5a, curve *1*). Microhardness of cleavages of the crystal part subjected to the action of a magnetic field (Fig. 5a, point *2*) decreased after keeping the crystal in a magnetic field at the same value as in the experiments of the first run. The influence of the magnetic field in the crystal bulk decreases with time. However, these changes have no component due to surface aging. Thus, the properties of potassium acid phthalate crystals kept in a magnetic field change not only at the surface but also in the crystal bulk.

Figure 5b shows the kinetics of microhardness changes after the application of the magnetic field. The $\Delta H(t)$ dependence corresponds to the difference in microhardness values (Figs. 4a, 4b, 5a; points *1* and *2*). It should be indicated that the $\Delta H$ value reflects the influence of the magnetic field on crystal microhardness. It is seen that the changes in microhardness in all the above experiments (Fig. 5b) in annealed and unannealed crystals vary with time almost in the same way and completely disappear after 6–7 days. Thus, this effect does not depend on the preliminary treatment of the sample. It should be remembered that, on the contrary, the magnetoplastic effect associated with the influence of a magnetic field on the state of the system of point defects in a crystal strongly depends on the preliminary thermal treatment of crystals. This is well seen in an example of alkali halide crystals [13]. Possibly, the structure of KAP crystals itself is changed under the action of a magnetic field.

As was mentioned above, potassium acid phthalate crystals are anisotropic; therefore, possible structural changes may be different at different faces. Therefore, in the next run of experiments, we performed indentation of the (100) face. To remove the defect layers formed due to sawing and to obtain smooth surfaces, we treated the crystals by the method considered above directly before the experiment. Then we broke each sample into two parts and, as in all the experiments performed, placed one part. into a 0.9 T magnetic field (**B** ∥ *z*) for 5 min, whereas the other part served as a reference sample. It turned out that microhardness in the (100) plane treated in a permanent magnetic field increased by 12–15% (within the experimental accuracy) and then remained constant with time (Fig. 6, curve *2*). Microhardness of the reference (100) surface increased after aging for four–five days by 12–15% (Fig. 6, curve *1*). Thus, the effect of magnetostimulated changes in microhardness of KAP crystals of the same orientation in a magnetic field has opposite signs on the perpendicular faces (Figs. 4a and 6). It should also be noted that the magnetic field blocked aging of both (010) and (100) surfaces of unannealed crystals.

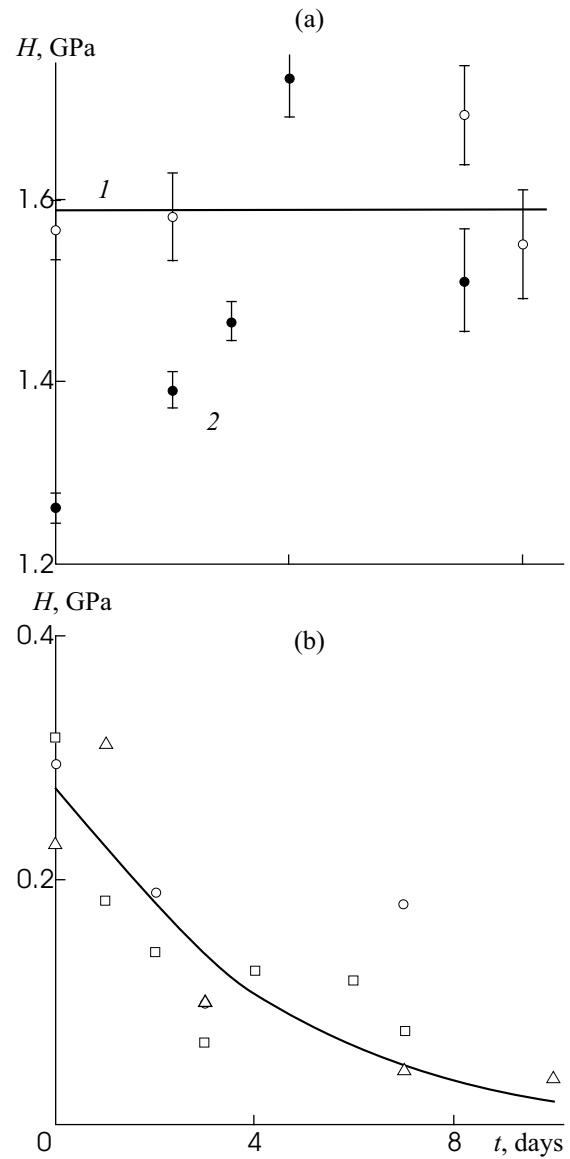

**Fig. 5.** (a) Microhardness and (b) differences in microhardness of the reference and magnetically treated samples in the (010) face of a potassium acid phthalate crystal as functions of time passed after 5-min treatment of samples in a 0.9 T magnetic field (**B** ∥ *z*); (a) each measurement was made (*1*) on an as-cleaved face of the reference sample and (*2*) on a sample treated in a magnetic field; (b) on unannealed samples (□), annealed samples (△), and on cleavages prepared from a bulky sample directly prior to each measurement (○).

Figure 7 shows a fragment of the model of a potassium acid phthalate crystal constructed on the basis of data from [18]. It is seen that the perpendicular faces of the crystal have different structures and face-forming elements. However, Figs. 4a and 6 show that aging of the (010) and (100) faces of unannealed crystals is blocked by a constant magnetic feld.

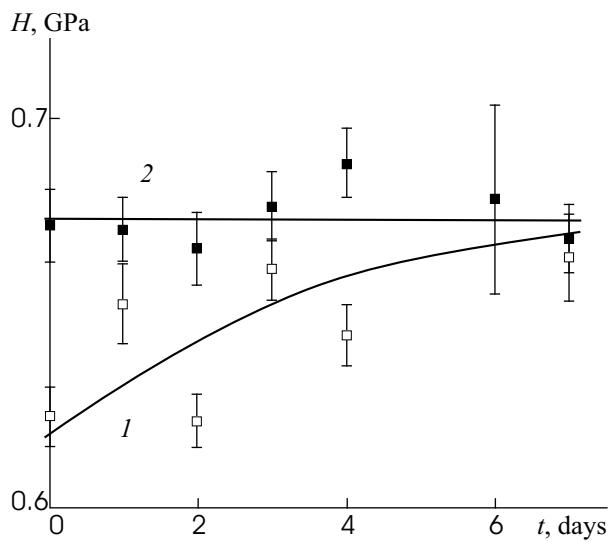

**Fig. 6.** Microhardness of the (100) face of a potassium acid phthalate crystal as a function of time passed after sample preparation: (*1*) for the reference sample, (*2*) for a sample treated for 5 min in a 0.9 T magnetic feld.

The effects observed in potassium acid phthalate crystals may hardly be explained by magnetostimulated changes in the local energy of interactions in the system dislocation–paramagnetic impurity center. First, the character of deformation described above shows that the main role is played by the displacements of the layers with respect to one another that are parallel to cleavage planes and not to generation and motion of dislocations as in alkali halide crystals. Second, unlike alkali halide crystals, in potassium acid phthalate crystals, the relative value of the effect does not depend on thermal treatment of crystals. Third, the change of microhardness after the equivalent action of a magnetic feld has opposite signs at the perpendicular faces of the crystal. This may readily be understood if one assumes that the crystal structure has changed. Then, because of anisotropy, a different reaction of the magnetically treated perpendicular faces to the indenter may have different signs. These changes may also hinder the removal of the interlayer water from the crystal, i.e., infuence the kinetic of surface aging. Of course, the above experimental data are still insuffcient for reliable consideration of structural changes in potassium acid phthalate crystals; however, it seems reasonable to analyze their structure.

The layers of $K^+$ ions are located in the parallel (010) planes at different heights, whereas cations form corrugated layers separated by double layers of anions. The hydrophobic parts (benzene rings) inside the layers "look" at one another and are linked by van der Waals interactions. Hydrophylic groups are attached to cationic layers and form the chains along the $z$ axis linked by intermolecular hydrogen bonds [18]. A fragment of such a chain is shown in the left-hand part of Fig. 7. Hydrogen bonds are depicted by dashed lines. It is possible to assume that, initially, some structural elements in a potassium acid phthalate crystal are in the metastable state but then relax. It seems that, in a magnetic feld, the conditions are created which stimulated their transition to an energetically more advantageous state. At this stage of the study, it is diffcult to say in what way a magnetic feld affects a KAP crystal and how it changes its structure. A structural element whose confgurational changes may be of a fuctuating nature may be a hydrogen bond, since the fuctuation may transform an O–H--O bond into an O--H–O bond [26], and a magnetic feld creates the conditions for such a transformation. Of course, we should not exclude possible transformation of some defect complexes related to structural elements under the action of a magnetic feld.

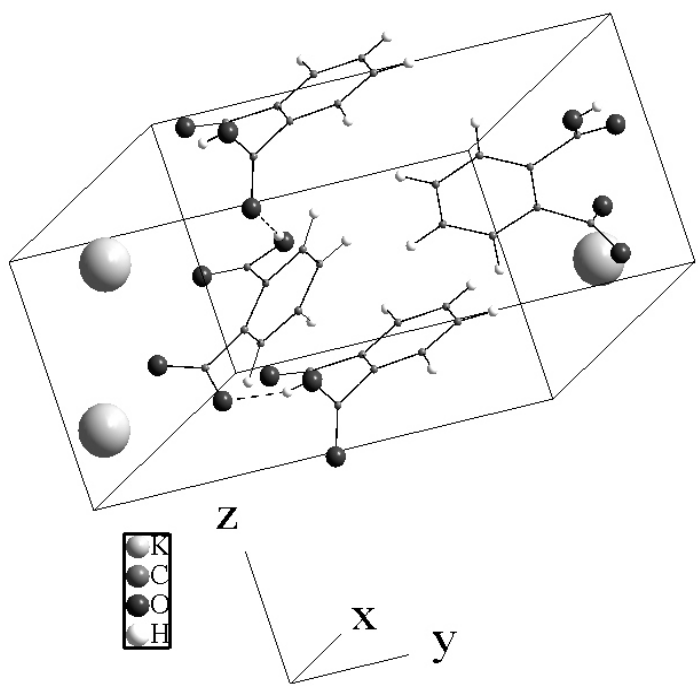

**Fig. 7.** Structural model of a potassium acid phthalate crystal and the unit cell and a chain of anions (solid line) linked by hydrogen bonds (dashed line).

The suggested interpretation of experimentally observed magnetostimulated change of microhardness in KAP crystals should be considered only as a working hypothesis. The verifcation of this hypothesis and the search for the explanations of such a pronounced effect of a magnetic feld on microhardness of potassium acid phthalate crystals require further studies with invocation of spectroscopic and X-ray diffraction methods.

## ACKNOWLEDGMENTS

This study was supported by the Russian Foundation for Basic research, project no. 03-02-17021.

We are also grateful to V.I. Alshits, A.I. Baranov, V.L. Berdinskiĭ, R.M. Zakalyukin, O.V. Klyavin, N.K. Moroz, E.A. Petrzhik, O.F. Pozdnyakov, N.L. Sizova, and N.G. Furmanova for fruitful discussions; R.B. Morgunov for the methodical help in crystal annealing; and Yu.G. Shvedenkova for measuring magnetic susceptibility.